\documentclass[a4paper,11pt]{article}
\usepackage{jheppub}
\usepackage{chessfss}
\pdfoutput=1
\pdfpageattr {/Group << /S /Transparency /I true /CS /DeviceRGB>>}
\usepackage{hyperref}
\hypersetup{colorlinks=true,linkcolor=blue,citecolor=blue}
\usepackage{graphicx,amsmath,amssymb,amsfonts,mathtools}
\usepackage{bm}
\usepackage{float}
\usepackage{subfig}
\usepackage{subfig}
\usepackage{scalefnt}
\usepackage{wrapfig}
\usepackage{fancybox}
\usepackage{caption}
\usepackage{tikz}
\usepackage{multirow}
\usetikzlibrary{arrows}
\tikzset{>=angle 60}
\tikzstyle{W}=[draw,circle,scale=.6]
\tikzstyle{B}=[draw,circle,fill=black,scale=.6]
\tikzstyle{H}=[draw,circle,fill=gray,scale=.6]
\tikzstyle{every picture}=[scale=.6,baseline=(current bounding box.south)]

\def\beq#1\eeq{\begin{align}#1\end{align}}

\newcommand{\be}{\begin{eqnarray}}
\newcommand{\ee}{\end{eqnarray}}
\newcommand{\bea}{\begin{eqnarray}}
\newcommand{\eea}{\end{eqnarray}}

\newcommand{\bn}{\begin{enumerate}}
\newcommand{\en}{\end{enumerate}}

\def\half{\frac{1}{2}}

\newtheorem{theo}{Theorem}

\title{Mirror Symmetry and Modularity}

\begin{document}
\author[\ast]{Babak Haghighat,}
\affiliation[\ast]{Yau Mathematical Sciences Center, Tsinghua University, Beijing, 100084, China}

\abstract{This is a review article on mirror symmetry and aspects of it related to the theory of modular forms. We describe this topic along its historical development and connect to some more recent results toward the end. The article is for publication in a special issue of ICCM Notices.}
\maketitle
%\flushbottom

\section{Introduction}
\label{sec:Intro}

The subject of mirror symmetry began with the discovery of an equivalence between two superconformal field theories \cite{Greene:1990ud} and was later shown to have a vast impact on enumerative geometry of Calabi-Yau manifolds \cite{Candelas:1990rm}. By now, the subject has evolved into a flourishing field of its own combining numerous topics in theoretical physics and mathematics. This review article aims to provide a short overview over aspects of mirror symmetry related to the theory of modular forms. In our exposition we will be following mainly the work of S.-T. Yau and collaborators who have provided important contributions to this sub-field of mirror symmetry.

To begin with, we note that mirror symmetry is the equivalence of two Calabi-Yau manifolds $M$ and $\widetilde{M}$ such that the complex structure of one manifold is exchanged with the K\"ahler structure of the other and vice versa. Denoting the complex structure parameters of one space (usually we take this space to be $M$) by $z_i$, $i=1,\ldots, h^{\textrm{dim}_{\mathbb{C}}M-1,1}(M)$, and the complexified K\"ahler parameters of the mirror manifold by $t_i$, $i=1,\ldots, h^{1,1}(\widetilde{M})$, there is a \textit{mirror map} $\vec{t}(\vec{z})$ relating the two
\begin{equation} \label{eq:mirrormap}
	\vec{t} \stackrel{\vec{t}(\vec{z})}{\longleftrightarrow} \vec{z}~.
\end{equation}
Another ingredient of mirror symmetry originating in physics is the so called \textit{topological string theory}. Here we subdivide two constructions, namely the topological $A$ model and the topological $B$ model. Roughly speaking, in the $A$ model one studies configurations of holomorphic maps from a genus $g$ Riemann surface to a Calabi-Yau manifold, whereas the $B$ model deals with constant maps. When the Riemann surface has no boundaries, one speaks of the \textit{closed} topological string, otherwise one is dealing with the \textit{open} topological string. Furthermore, one shows that correlation functions of the $A$ model only depend on the symplectic or K\"ahler structure of the Calabi-Yau, whereas those of the $B$ model only depend on complex structures. One then defines partition functions of these theories denoted by $Z^{top}_A$ and $Z^{top}_B$ which depend on the moduli and the topological string coupling constant $\lambda$ which keeps track of the genus of the Riemann surface appearing in the definition of the models described above. The mirror map (\ref{eq:mirrormap}) then relates the two partition functions \cite{Bershadsky:1993cx}. 

Mathematically, the topological $A$ model partition function encodes Gromov-Witten generating functions.  To extract them, we have to rewrite the closed topological string partition function in terms of free energies $F^g$ as follows
\begin{equation}
	Z^{top}(\vec{t},\lambda) = e^{\sum_{g=0}^{\infty} \lambda^{2g-2} F^g(\vec{t})},
\end{equation}
where the $F^g$ themselves admit at large K\"ahler structure in the moduli space an expansion of the form\footnote{We are focusing here only on the instanton contributions of string theory and neglect perturbative terms.}
\begin{equation}
	F^g = \sum_{\vec{d}} N^g_{\vec{d}} \prod_{i=1}^{h^{1,1}(\widetilde{M})} q_i^{d_i}, \quad q_i = e^{2\pi i t_i}.
\end{equation}
In the above formula, the numbers $N^g_{\vec{d}}$ are rational numbers known as Gromov-Witten invariants associated to holomorphic curves of genus $g$ and degree $d_i$, $i=1, \ldots, h^{1,1}(\widetilde{M},\mathbb{Z})$, with respect to generators of the K\"ahler cone. In the following, we want to understand under what circumstances the functions $F^g$ and the mirror map can be expanded in terms of modular forms and functions. By modular we mean here modularity with respect to the group $SL(2,\mathbb{Z})$ and its subgroups, although in Section \ref{sec:ADE} we will be encountering a case where the modular group is $Sp(2,\mathbb{Z})$. That is, we will be identifying a distinguished modulus $t$ (or several distinguished moduli in the case of $Sp(2,\mathbb{Z})$) whose modular transformations are symmetries of the topological string free energies. The results we review are mainly derived in the context of the genus $0$ free energies $F^0$ but most of the emerging structure can be carried over to the higher genus case. Such modularity properties are very important from the computational point of view, as they allow to solve for the $F^g$ much more efficiently due to the finite basis properties of rings of modular forms. Also, results in these directions have great importance for number theory as the free energies are conjectured to admit expansions with integral coefficients \cite{Gopakumar:1998jq}.

The organization of this article is as follows. We will start along the historical path modularity showed up in mirror symmetry, namely in terms of solutions of differential equations. Such differential equations are known as Picard-Fuchs equations and they have played a prominent role in the development of mirror symmetry. We will be reviewing the cases of the elliptic curve, the K3 surface and the Quintic Calabi-Yau manifold in Sections \ref{sec:ellcurve}, \ref{sec:K3}, \ref{sec:Quintic}, before proceeding to the more complex cases of elliptic fibrations in Section \ref{sec:fibrations}. The structure encountered in the study of elliptic fibrations then motivates us to shift gears and take a new look at the elliptic curve in Section \ref{sec:newellcurve} by studying it from the perspective of homological mirror symmetry of Kontsevich \cite{Kontsevich:1994dn}. We then proceed in Section \ref{sec:ADE} to the case of a non-compact Calabi-Yau threefold admitting two elliptic fibrations and connect to previous results.

\section{Elliptic Curve}
\label{sec:ellcurve}
Let us first review mirror symmetry for elliptic curves where we will be following reference \cite{Klemm:1994wn}. Here $M = E_{\tau}$ is an elliptic curve with modular parameter $\tau$ and $\widetilde{M} = \widetilde{E}^{t}$ is the torus $\mathbb{R}^2/(\mathbb{Z}\oplus \mathbb{Z})$ with complexified K\"ahler parameter $t = iA + b$, where $A$ is the area and $b$ defines a class in $H^2(\widetilde{M};\mathbb{R})/H^2(\widetilde{M};\mathbb{Z})$. The mirror map is 
\begin{equation}
	t \leftrightarrow \tau
\end{equation}
To be more specific, consider an explicit realization of $E_{\tau}$ as the family of cubic curves in $\mathbb{P}^2$:
\begin{equation}
	X_s : x_1^3 + x_2^3 + x_3^3 - s x_1 x_2 x_3 = 0.
\end{equation}
Using Nagell's algorithm \cite{EllipticCurveHandBook} one can bring the above curve into Weierstrass form:
\begin{equation}
	y^2 = 4 x^3 - g_2 x - g_3,
\end{equation}
where 
\begin{eqnarray}
	g_2 & = & 72 s + \frac{1}{3}s^4 \nonumber \\
	g_3 & = & 432 + 40s^3 - \frac{2}{27}s^6.
\end{eqnarray}
From this form one can read off the discriminant of the curve
\begin{equation}
	\Delta = 4 g_2^3 - 27 g_3^2 = 256 (-27 + s^3)^3.
\end{equation}
Next, we consider the period of the holomorphic 1-form $\frac{dx}{y}$ along a homology cycle $\Gamma$:
\begin{equation}
	\omega_{\Gamma} = \int_{\Gamma} \frac{dx}{y}.
\end{equation}
One can then show that as a function of $s$, $\omega_{\Gamma}$ satisfies the second order ODE given by
\begin{equation} \label{eq:perioddiffeq}
	\frac{d^2 \omega_{\Gamma}}{ds^2} + a_1(s) \frac{d\omega_{\Gamma}}{ds} + a_0(s) \omega_{\Gamma} = 0,
\end{equation}
where 
\begin{eqnarray}
	a_1 & = & - \frac{d}{ds}\log\left(\frac{3}{2\Delta}(2g_2\frac{dg_3}{ds}-3 \frac{dg_2}{ds}g_3)\right)\nonumber \\
	a_0 & = & \frac{1}{12}a_1 \frac{d}{ds}\log \Delta + \frac{1}{12} \frac{d^2}{ds^2}\log \Delta - \frac{1}{16 \Delta}(g_2 \frac{dg_2}{ds}^2 - 12 \frac{dg_3}{ds}^2).
\end{eqnarray}
Performing the change of coordinates $s \rightarrow z = s^{-3}$, equation (\ref{eq:perioddiffeq}) transforms into the following hypergeometric equation 
\begin{equation}
	(\theta^2 - 3z(3\theta + 2)(3 \theta + 1)) \omega_{\Gamma} = 0,
\end{equation}
where $\theta = z \partial_z$. This equation is characterized by having regular singularities at $z = 0, 1/3^3, \infty$. The mirror coordinate $t$ is then defined by taking the ratio of two periods $\omega_{\Gamma}$, $\omega_{\Gamma'}$. Inversion of the function $t(z)$ then determines the \textit{mirror map}. Up to now we have solely been dealing with an elliptic curve realized as a hypersurface in $\mathbb{P}^2$. More generally, one can consider realizations as hypersurfaces in weighted projective spaces $\mathbb{P}^2(1,2,3)$ and $\mathbb{P}^2(1,1,2)$. The corresponding hypersurface and Picard-Fuchs equations are summarized in Table \ref{tab:PFtab}.
\begin{table}[h]
\centering
\begin{tabular}{|c|cc|}
\hline
\textrm{ambient space} & \textrm{constraint} & \textrm{Picard-Fuchs operator} \\
\hline 
$\mathbb{P}^2(1,1,1)$ & $x_1^3 + x_2^3 + x_3^3 - z^{-1/3} x_1 x_2 x_3 = 0$ & $\theta^2 - 3z(3\theta + 2)(3\theta + 1)$  \\
$\mathbb{P}^2(1,1,2)$ & $x_1^4 + x_2^4 + x_3^2 - z^{-1/4} x_1 x_2 x_3 = 0$ & $\theta^2 - 4z(4\theta + 3)(4\theta + 1)$ \\
$\mathbb{P}^2(1,2,3)$ & $x_1^6 + x_2^3 + x_3^2 - z^{-1/6} x_1 x_2 x_3 = 0$ & $\theta^2 - 12 z (6\theta + 5)(6\theta + 1)$\\
\hline
\end{tabular}
\caption{Picard-Fuchs equations for elliptic curves embedded in weighted projective spaces.}
\label{tab:PFtab}
\end{table}
There exists a more unified approach for dealing with the Picard-Fuchs operator in the case of elliptic curves which we shall briefly describe here. The first step is to perform the following change of coordinates $s \rightarrow J = \frac{g_2^3}{\Delta}$, and write $\omega_{\Gamma}$ as $\sqrt{\frac{g_2}{g_3}}\Omega_{\Gamma}$. Then equation (\ref{eq:perioddiffeq}) becomes
\begin{equation} \label{eq:universalpf}
	\frac{d^2\Omega_{\Gamma}}{dJ^2} + \frac{1}{J} \frac{d \Omega_{\Gamma}}{dJ} + \frac{31 J -4}{144J^2 (1-J)^2} \Omega_{\Gamma} = 0.
\end{equation}
As this equation is derived without the use of the explicit form of $g_2$ and $g_3$, it is universally applicable to elliptic curves even beyond the case of the cubic in $\mathbb{P}^2$. This means that if we start from any other model for an elliptic curve, we will still arrive at the same equation (\ref{eq:universalpf}). Solving this equation and taking again ratios of two periods we obtain the mirror coordinate $t$ as a function of $J$. The differential equation governing the dependence between the two variables is the Schwarzian equation:
\begin{equation}
	\{t,J\} = 2 \left(\frac{3}{16(1-J)^2} + \frac{2}{9J^2} + \frac{23}{144J(1-J)}\right).
\end{equation}
Here $\{z,x\}$ denotes the Schwarzian derivative $\frac{z'''}{z'} - \frac{3}{2} \left(\frac{z''}{z'}\right)^2$. Inverting the function $t(J)$ we obtain the mirror map $J(t)$. Performing all steps one arrives, with a suitable choice of period ratio $t$, at the following well-known expression for the $J$-function of elliptic curves
\begin{equation}
	J(q) = \frac{1}{1728}(q^{-1} + 744 + 196884 q + 21493760 q^2 + \ldots), \quad q=e^{2\pi i t}.
\end{equation}
This form makes the invariance under modular $SL(2,\mathbb{Z})$ transformations of $t$ manifest.
In order to obtain the individual mirror maps for various elliptic curves, one has to substitute in the function $t(J)$ the function $J \rightarrow J(z)$ with $z$ being the complex structure parameter of the curve. This can be done by using the explicit form of $g_2$ and $\Delta$. We summarize the functional dependence on $z$ for elliptic curves which are realized as hypersurfaces in weighted projective spaces $\mathbb{P}^2(1,1,1)$, $\mathbb{P}^2(1,1,2)$ and $\mathbb{P}^2(1,2,3)$ in Table \ref{tab:Jtab}.
\begin{table}[h]
\centering
\begin{tabular}{|c|cc|}
\hline 
\textrm{ambient space} & \textrm{constraint} & $1728 J(z)$ \\
\hline
$\mathbb{P}^2(1,1,1)$ & $x_1^3 + x_2^3 + x_3^3 - z^{-1/3} x_1 x_2 x_3 = 0$ & $\frac{(1+216 z)^3}{z(1-27z)^3}$  \\
$\mathbb{P}^2(1,1,2)$ & $x_1^4 + x_2^4 + x_3^2 - z^{-1/4} x_1 x_2 x_3 = 0$ & $\frac{(1+192z)^3}{z(1-64z)^2}$ \\
$\mathbb{P}^2(1,2,3)$ & $x_1^6 + x_2^3 + x_3^2 - z^{-1/6} x_1 x_2 x_3 = 0$ & $\frac{1}{z(1-432z)}$\\
\hline
\end{tabular}
\caption{The $j$-functions for various elliptic curves embedding in weighted projective spaces.}
\label{tab:Jtab}
\end{table}

\section{K3 surface}
\label{sec:K3}

Let us consider the following one-parameter family of quartic hypersurfaces in $\mathbb{P}^3$:
\begin{equation}
	X_s : W_s(x_1,x_2,x_3,x_4) = x_1^4 + x_2^4 + x_3^4 +x_4^4 - s x_1 x_2 x_3 x_4 = 0.
\end{equation}
The period of a holomorphic 2-form along a homology 2-cycle $\Gamma_i$ in $X_s$ is given by
\begin{equation}
	\omega_i(s) = \int_{\gamma} \int_{\Gamma_i} \frac{\mu}{\sum_{i=1}^4 x_i^4 - s\prod_{i=1}^4 x_i},
\end{equation}
where $\gamma$ is a small cycle in $\mathbb{P}^{4-1}$, $\Gamma_i \in H_2(X_s)$ and the measure is
\begin{equation}
	\mu = \sum_i (-1)^i x_i dx_1 \wedge \ldots \widehat{dx_i} \ldots \wedge dx_4.
\end{equation}
The corresponding Picard-Fuchs operator annihilating the periods $\omega_i$ is of the form \cite{Lian:1995js}
\begin{equation} \label{eq:deg4K3}
	\mathcal{L} = \theta^3 - 8z(1+2\theta)(1+4\theta)(3+4 \theta).
\end{equation}
By the Frobenius method, a basis of solutions is given by
\begin{equation}
	\omega_i(z) = \left(\frac{1}{2\pi i} \frac{\partial}{\partial \rho}\right)^i \sum_{k\geq 0}\left. \frac{\Gamma(4(k+\rho)+1)}{\Gamma(k+\rho+1)^4}z^{k+\rho}\right|_{\rho=0}, \quad i=0,\ldots, 2.
\end{equation}
If we take $q = e^{2\pi i t}$, then the mirror map is given by the $q$-series obtained by inverting the relation
\begin{equation}
	q = \exp\left(\frac{\omega_1(z)}{\omega_2(z)}\right).
\end{equation}
Now one can ask whether there are corresponding Schwarzian equations and modularity properties similar to the elliptic curve case. It turns out that indeed such properties do exist, as shown in \cite{Lian:1995js} and as we will review in the following for the general setting. For a general one-parameter family of K3 surfaces the corresponding Picard-Fuchs operator is given by
\begin{equation} \label{eq:K3pf}
	\mathcal{L} = \theta^3 - \lambda z (\theta + 1/2)(\theta + 1/2 + \nu)(\theta + 1/2 - \nu),
\end{equation}
where $\lambda$ and $\nu$ are arbitrary complex numbers\footnote{In fact, in the practical situation one comes across, these variables are always integers.}. From the classical theory of Schwarzian equations, it follows that the mirror map $z(q)$ corresponding to the Picard-Fuchs operator (\ref{eq:K3pf}) is a solution to
\begin{equation}
	2 Q z'^2 + \{z,t\} = 0,
\end{equation}
where 
\begin{equation}
	Q = \frac{1 + (-\frac{5}{4} + \nu^2)\lambda z + (1-\nu)(1+\nu)\lambda^2 z^2}{4 x^2 (1-\lambda x)^2}.
\end{equation}
On the other hand, it is known that the unique power series solution with leading term $1 + \mathcal{O}(z)$ to the differential operator $\mathcal{L}$ is the hypergeometric function $\prescript{}{3}{F_2(\half,\half + \nu, \half - \nu;1,1;\lambda z)}$. Furthermore, one can show that
\begin{equation}
	\mathcal{L} \frac{z'}{z(1-\lambda z)^{1/2}} = (1-\lambda z)^{1/2} z^2 z'^{-2} \frac{d}{dt}(2 Q z'^2 + \{z,t\}),
\end{equation}
and thus vanishes due to the Schwarzian equation. Combining all statements, one arrives at the following identity \cite{Lian:1995js}
\begin{equation} \label{eq:K3modularity}
	\prescript{}{3}{F_2(\half,\half+\nu,\half-\nu;1;1;\lambda z(q))^2} = \frac{z'^2}{z^2 (1-\lambda z)}.
\end{equation}
Coming back to our example (\ref{eq:K3pf}), the assignments are $(\lambda,\nu) = (2^8,2^{-2})$ and the left-hand side of (\ref{eq:K3modularity}) is a weight $4$ modular form of the genus zero group $\Gamma_0(2)+$, while $z(t)$ is its respective Hauptmodul. Similarly, for the cases $(\lambda,\nu) = (2^2 3^3,2^{-1} 3^{-1})$ and $(\lambda, \nu) = (2^6,0)$ one obtains the Hauptmoduls for the groups $\Gamma_0(3)+$ and $\Gamma_0(4)+$ respectively.

\section{Quintic threefold}
\label{sec:Quintic}

Mirror symmetry for the quintic hypersurface in $\mathbb{P}^4$ was first studied in the famous work \cite{Candelas:1990rm}. We will review certain features of the construction here. Let us start with the hypersurface constraint which is given by the following equation:
\begin{equation}
	X_{\psi} = x_1^5 + x_2^5 + x_3^5 + x_4^5 + x_5^5 - 5 \psi^{\frac{1}{5}} x_1 x_2 x_3 x_4 x_5 = 0.
\end{equation}
Performing the variable change $z = \frac{1}{5^5 \psi}$, one can show that the corresponding Picard-Fuchs operator annihilating the periods of this Calabi-Yau is given by
\begin{equation}
	\mathcal{L} = \theta^4 - 5 z (5\theta +1)(5\theta+2)(5\theta+3)(5\theta+4).
\end{equation}
A basis of solutions is specified by
\begin{equation}
	\omega_i(z) = \left(\frac{1}{2\pi i} \frac{\partial}{\partial \rho}\right)^i \sum_{k\geq 0}\left. \frac{\Gamma(5(k+\rho)+1)}{\Gamma(k+\rho+1)^5}z^{k+\rho}\right|_{\rho=0}, \quad i=0,\ldots, 3.
\end{equation}
The mirror map is defined as follows
\begin{equation}
	2\pi i t(\psi) = \int_{\mathcal{C}} (iJ + B) = \frac{\omega_1}{\omega_0} = -\log(5^5 \psi) + \frac{154}{625 \psi} + \frac{28713}{390625\psi^2} + \ldots
\end{equation}
and its inversion is given by the expansion
\begin{equation}
	\frac{1}{z} 5^5 \psi = \frac{1}{q} + 770 + 421375 q + 274007500 q^2 + \ldots, \quad q = e^{2\pi i t}.
\end{equation}
The special geometry prepotential $F^0$ of the Calabi-Yau three-fold is then determined by a particular linear combination of the periods $\omega_i$ involving the triply-logarithmic solution $\omega_3$ as for example reviewed in \cite{Huang:2006hq}. Inserting the mirror map then gives the instanton expansion of the prepotential as follows
\begin{equation}
	F^0 = - \frac{\kappa}{3!}t^3 - \frac{a}{2}t^2 + ct + \frac{e}{2} + f_{inst}(q),
\end{equation}
In the above, $\kappa$ denotes the triple-intersection number of the Fermat Quintic, while $a$, $c$ and $e$ are special topolotical constants of the manifold. We have
\begin{eqnarray}
	c & = & \frac{1}{24} \int_M c_2(T_M)\wedge H \nonumber \\
	e & = & \frac{\zeta(3)}{(2\pi )^3i} \int_M c_3(T_M) \nonumber \\
	a & = & \half \int_M i_* c_1(H) \wedge H,
\end{eqnarray}
where in the above, $H$ denotes the hyperplane class of the Fermat Quintic and $i$ is the embedding map. Similar results can be obtained for a wide class of hypersurfaces and complete intersections in weighted projective space \cite{Hosono:1993qy,Hosono:1994ax}. 

What is lacking in the case of the quintic as compared to the previous cases is a modularity structure for the periods. This is indeed not clear at the moment and remains one of the central questions in the realm of the Quintic mirror symmetry. It would be desirable to show how such modularity properties can arise and to what extent they can be used to solve mirror symmetry completely. For some progress along these directions see \cite{Alim:2014dea}. However, one does find modular properties of periods for another class of compact Calabi-Yau threefolds, namely for the class of elliptic fibrations to which we turn our attention in the following section.

\section{Elliptic Fibrations}
\label{sec:fibrations}

In this section we want to focus on novel features of periods and mirror maps in the case of elliptic fibrations giving rise to modular properties. We shall review these properties in the light of the work \cite{Haghighat:2015qdq}. This paper studies Calabi-Yau $n$-folds $\widetilde{X}$ which are elliptic fibrations over $\mathbb{P}^{n-1}$. The corresponding Picard-Fuchs operators are given by
\begin{eqnarray} 
	L_1 & := & - n \cdot \theta_1 \theta_2 + \theta_1^2 - a_0 z_1(\theta_1 + a_1)(\theta_1 + a_2) = 0, \nonumber \\
	L_2 & := & \theta_2^n - (-1)^n z_2(n\cdot \theta_2 - \theta_1)(n \cdot \theta_2 - \theta_1 + 1) \cdots (n \cdot \theta_2 - \theta_1 + n -1) = 0, \label{eq:PFssystem}
\end{eqnarray}
where $n$, $a_0$, $a_1$, $a_2$ are parameters of the system. The relevant cases to String Theory are the cases $n=3,4$ and $(a_0,a_1,a_2)$ as in Table \ref{tab:modulargroups}.
\begin{table}
\centering
\begin{tabular}{|c|c|c|}
\hline
$(a_0,a_1,a_2)$  & Group & Modular forms \\ \hline
$(432,5/6,1/6)$ & $SL(2,\mathbb{Z})$ & $E_4(\tau),E_6(\tau)$ \\ \hline
$(64,3/4,1/4)$  & $\Gamma_0(2)$  & $E_2(\tau)-2E_2(2\tau),\ E_4(\tau)$ \\ \hline
$(27,2/3,1/3)$  &  $\Gamma_0(3)$ &  $E_2(\tau)-3E_2(3\tau),\ E_4(\tau), E_6(\tau)$  \\ \hline
$(16,1/2,1/2)$  &   $\Gamma(2)$ &  $\theta_2^4,\theta_3^4$  \\
\hline
\end{tabular}
\caption{Modular groups}
\label{tab:modulargroups}
\end{table}
If we define $\mathcal{L}_n \subset [z_1,z_2,\theta_1,\theta_2]$, with $\theta_i = z_i \partial_{z_i}$, to be the differential left ideal generated by the operators $L_1$ and $L_2$, the $\mathcal{L}_n$ annihilates the periods of a $(n,0)$-form $\omega^{(n,0)}$ of the mirror Calabi-Yau manifold $X_z$. The system $\mathcal{L}_n$ has one holomorphic solution $\Pi^0 = \mathcal{O}(1)$ and logarithmic solutions $\Pi^a = \Pi^0 \log(z_a) + \mathcal{O}(1)$, $a=1,2$. 

One central ingredient in the observations of \cite{Haghighat:2015qdq} is the field $\textrm{M}_n$ of differential Calabi-Yau modular forms which is the field extension of $\mathbb{C}$ generated by
\begin{align} \label{eq:Mn}
	z_1, z_2, \quad \theta_1^i \theta_2^j \Pi^0, \quad \theta_1^i \theta_2^j \left(\Pi^0 \theta_a \Pi^b - \Pi^b \theta_a \Pi^0 \right) \nonumber \\
	a,b = 1,2, \ldots, \textrm{h} := h^{12}(X_z) = 2, \quad i,j \in \mathbb{N}_0.
\end{align}
Elements of $\textrm{M}_n$ are by definition rational functions in (\ref{eq:Mn}) with coefficients in $\mathbb{C}$. It can be argued that all topological string free energies $F^g$ are completely expressible in terms of elements of $\textrm{M}_n$. The modular expressions of the elements of $\textrm{M}_n$ are obtained after inserting the mirror map $(\tau_1,\tau_2) = (\frac{\Pi^1}{\Pi^0},\frac{\Pi^2}{\Pi^0})$ or using the $(q_1,q_2) = (e^{\tau_1},e^{\tau_2})$ coordinates. In the following we will use the same symbol for an element $f(x)$ of $\textrm{M}_n$ when working with different coordinate systems $x=(z_1,z_2), (\tau_1,\tau_2)$ or $(q_1,q_2)$. Then the main result of \cite{Haghighat:2015qdq} can be stated as
\begin{theo}
\label{maintheo}
 Let $f(q_1,q_2)\in \textrm{M}_n$ and assume that it is of the form
 $$
 f=f_0(q_1)+f_1(q_1)(q_2q_1^{\frac{n}{2}})+\cdots +f_i(q_1)(q_2 q_1^{\frac{n}{2}})^i+\cdots
 $$
 where $q_1^{\frac{n}{2}}=e^{\tau_1\cdot \frac{n}{2}}$.
 Then for arbitrary $n$ and $(a_0, a_1,a_2)$ as in Table  
 \ref{tab:modulargroups}  all $f_i(e^{\tau_1})$'s are algebraic over the field of meromorphic 
 quasi-modular forms on the upper half plane $\tau_1\in \mathbb{H}$ 
 for the  subgroup of $SL(2,\mathbb{Z})$ listed in the same table. % and with poles at the cusp $\tau_1=i\infty$. 
 \end{theo}
In the $SL(2,\mathbb{Z})$ case one finds \cite{Haghighat:2015qdq} (see also \cite{Klemm:2012sx,Alim:2012ss} for the threefold case) that for Calabi-Yau $n$-folds all $f_i(q_1)$ are of the form
\begin{equation} \label{eq:polform}
	f_i(q_1) = \frac{P(E_2,E_4,E_6)}{\eta^{12 n i}},
\end{equation}
where $P(E_2,E_4,E_6)$ denotes polynomials in the Eisenstein series $E_2$, $E_4$ and $E_6$. This functional form is a subset of the allowed $f_i$ appearing in the above theorem. It would be very interesting to analyze the proof of Theorem \ref{maintheo} and obtain a finer statement which shows why the particular form (\ref{eq:polform}) appears. The central idea of the proof is the observation that in the limit $z_2 \rightarrow 0$ the Picard-Fuchs system (\ref{eq:PFssystem}) reduces to an ideal generated by the operator
\begin{equation} \label{eq:reducedPFsystem}
	L := \theta_1^2 - a_0 z (\theta_1 + a_1)(\theta_1 + a_2).
\end{equation}
One sees here that the Picard-Fuchs operator $L$ is of the same form as the ones for the elliptic curve discussed in Section \ref{sec:ellcurve}. Therefore, solutions of (\ref{eq:reducedPFsystem}) are the well-known Gauss hypergeometric functions 
$F(a_1,a_2,1|z)$ and their derivatives. Inserting the mirror map
\begin{equation}
	z \rightarrow z(\tau) = \frac{1}{864}(1 - \sqrt{1- 1728/J}),
\end{equation}
then gives expressions in terms modular forms through the following replacements
\begin{equation}
	F(z) \rightarrow F(z(\tau)) = (E_4)^{\frac{1}{4}}, \quad \theta F(z) \rightarrow \theta F(z(\tau)) = \frac{E_4^{1/4}(E_2 E_4 - E_6)}{6 (E_4^{3/2} + E_6)}.
\end{equation}
We will give an example of this in the case of Calabi-Yau fourfolds in the following. Mirror symmetry of elliptic fourfolds was first studied extensively in the seminal paper \cite{Klemm:1996ts} where Picard-Fuchs equations and genus $0$ Gromov-Witten invariants for a wide class of elliptic Calabi-Yau fourfolds were derived. However, modularity properties for Gromov-Witten generating functions have appeared only recently \cite{Haghighat:2015qdq}. The main example of \cite{Haghighat:2015qdq} is an elliptic fibration over $\mathbb{P}^3$ which is the resolution of the degree $24$ orbifold hypersurface 
\begin{equation}
	\widetilde{X} \subset \mathbb{P}^5(1,1,1,1,8,12)
\end{equation}
in weighted projective space. The resolution has the following non-vanishing Hodge numbers \cite{Klemm:2007in}
\begin{equation}
	h_{0,0} = h_{4,0} = 1, \quad h_{11} =2, \quad h_{31} = 3878, \quad h_{22} = 15564.
\end{equation}
$H^{1,1}(\widetilde{X})$ is generated by the two elements $J_1$ and $J_2$ which are Poincare dual to the elliptic fiber $E$ and base $B$ respectively in the notation of \cite{Klemm:1996ts}. The genus $0$ Gromov-Witten generating functions in the case of fourfolds depend further on a choice of four-cycle. In our case we have the following basis of $H_V^{2,2}$:
\begin{equation}
	\gamma_1 := J_1^2, \quad \gamma_2 := \frac{1}{17}(4 J_1^2 + J_1 J_2).
\end{equation}
In this basis one then has the following solutions for the Gromov-Witten generating functions
\begin{eqnarray}
	F^0(\gamma_1) & = & q_2 \left(\frac{q_1^2}{\eta^{48}}\right)\left[\frac{5}{18}E_4 E_6 (35 E_4^3 + 37 E_6^2)\right] + \mathcal{O}(q_2^2), \nonumber 	\\
	F^0(\gamma_2) & = & 1 + q_2 \left(\frac{q_1^2}{\eta^{48}}\right)]\left[\frac{5}{10368}(10321 E_4^6 + 1680 (-24 + E_2)E_4^4 E_6) + 59182 E_4^3 E_6^2\right. \nonumber \\
	~ & ~ & \left. + 1776(-24 + E_2) E_4 E_6^3 + 9985 E_6^4\right] + \mathcal{O}(q_2^2) .
\end{eqnarray}
Given the importance of the role of mirror symmetry for the elliptic curve in the derivation of the main Theorem \ref{maintheo}, one might wonder whether a more thorough investigation of the elliptic curve will reveal further key aspects of mirror symmetry for elliptically fibred Calabi-Yau manifolds. We shall pursue this path of thought in the coming section by taking a new look at the elliptic curve.

\subsection{A new look at the elliptic curve}
\label{sec:newellcurve}

In this section we will outline a new direction by examining categorical mirror symmetry for the elliptic curve as described by Zaslow and Polishchuk in \cite{Polishchuk:1998db}. Categorical or homological mirror symmetry was first conjectured by Kontsevich in the seminal paper \cite{Kontsevich:1994dn}. Kontsevich's conjecture relates two mirror manifolds $M$ and $\widetilde{M}$ by an equivalence of categories. On the one side, one considers the bounded derived category of coherent sheaves on $M$, denoted by $\mathcal{D}^b(M)$, and on the other side the derived category of a suitable enlargement of Fukaya's $A^{\infty}$-category $\mathcal{F}(\widetilde{M})$ of minimal Lagrangian submanifolds on $\widetilde{M}$: $\mathcal{D}^b(M) \cong \mathcal{D}^b(\mathcal{F}(\widetilde{M}))$. The work \cite{Kontsevich:1994dn} had been done without the knowledge of D-branes. But it can be understood most intuitively by using the notion and dualities of D-branes as shown in \cite{Strominger:1996it}. On a Calabi-Yau three-fold in Type IIB string theory, D-branes correspond to minimal Lagrangian submanifolds with flat $U(n)$ gauge bundles. This side corresponds to the Fukaya side. On the other side, the derived category $\mathcal{D}^b(M)$ corresponds to even-dimensional D-branes of Type IIA string theory wrapping holomorphic submanifolds of the mirror Calabi-Yau. The conjecture of \cite{Strominger:1996it} then relates these two sides beautifully and states that D-branes are in fact responsible for mirror symmetry.

In the following exposition we will not need all the mathematical details of the two categories $\mathcal{D}^b(M)$ and $\mathcal{D}^b(\mathcal{F}(\widetilde{M}))$ and will only focus on the most elementary properties in the case where $M$ is an elliptic curve. Given the definitions in Section \ref{sec:ellcurve}, we have to address the equivalence\footnote{As explained in \cite{Polishchuk:1998db}, in the case where $M$ is the elliptic curve it is safe to replace $\mathcal{D}^b(\mathcal{F}(\widetilde{M}))$ by $\mathcal{F}^0(\widetilde{M})$.}
\begin{equation}
	\mathcal{D}^b(E_{\tau}) \cong \mathcal{F}^0(\widetilde{E}^{\tau}).
\end{equation}
Let us begin with the right-hand side of the above equivalence, namely the category $\mathcal{F}$. The objects of this category are special Lagrangian submanifolds $\mathcal{L}_i$ endowed with flat bundles $\mathcal{E}_i$ with monodromies having eigenvalues of unit modulus.\footnote{Recall that a special Lagrangian submanifold $L$ is a middle-dimensional submanifold satisfying $\left. \omega \right|_L = 0$ and $\left.\textrm{Im}(\Omega)\right|_L = 0$ where $\omega$ and $\Omega$ are the K\"ahler form and the holomorphic three-form of the Calabi-Yau $\widetilde{M}$ respectively.}. We hence denote an object $\mathcal{U}_i$ by a pair
\begin{equation}
	\mathcal{U}_i = (\mathcal{L}_i,\mathcal{E}_i).
\end{equation}
The morphisms are open strings stretching between two branes. Mathematically, they are elements of the space $\textrm{Hom}(\mathcal{U}_i,\mathcal{U}_j)$ defined as
\begin{equation}
	\textrm{Hom}(\mathcal{U}_i,\mathcal{U}_j) = \mathbb{C}^{\#\{\mathcal{L}_i \cap \mathcal{L}_j\}} \otimes \textrm{Hom}(\mathcal{E}_i,\mathcal{E}_j).
\end{equation}
Furthermore, to each open string state we assign an integer denoting the corresponding ghost number, thus making the above space $\mathbb{Z}$-graded. To summarize, for any two objects $X$ and $Y$ inside $\textrm{Ob}(\mathcal{F})$, the morphisms are given by a $\mathbb{Z}$-graded abelian group $\textrm{Hom}(X,Y)$. The joining of two adjacent strings is described by a composition of maps
\begin{equation}
	m_k : \textrm{Hom}(X_1,X_2) \otimes \textrm{Hom}(X_2,X_3) \otimes \ldots \textrm{Hom}(X_k,X_{k+1}) \rightarrow \textrm{Hom}(X_1,X_{k+1}),
\end{equation}
$k \geq 1$, of degree $2-k$, satisfying the following \textit{associativity} condition
\begin{equation}
	\sum_{r=1}^n \sum_{s=1}^{n-r+1} (-1)^{\epsilon} m_{n-r+1}(a_1 \otimes \ldots \otimes a_{s-1} \otimes m_r(a_s \otimes \ldots \otimes a_{a+r-1})\otimes a_{s+r} \otimes \ldots \otimes a_n) = 0
\end{equation}
for all $n \geq 1$, where $\epsilon = (r+1)s + r(n + \sum_{j=1}^{s-1} \textrm{deg}(a_j))$. The $A^{\infty}$ structure on Fukaya's category is given by summing over holomorphic maps from the disc $D^2$, which take the components of the boundary $S^1 = \partial D^2$ to the special  Lagrangian objects. These maps can be thought of as world-sheets of open topological strings ending on the Lagrangian branes. An element $u_j$ of $\textrm{Hom}(\mathcal{U}_j, \mathcal{U}_{j+1})$ is represented by a pair
\begin{equation}
	u_j = t_j \cdot a_j,
\end{equation}
where $a_j \in \mathcal{L}_j \cap \mathcal{L}_{j+1}$, and $t_j$ is a matrix in $\textrm{Hom}(\left.\mathcal{E}_j\right|_{a_j},\left.\mathcal{E}_{j+1}\right|_{a_j})$. Then we have
\begin{equation} \label{eq:gluing}
	m_k(u_1 \otimes \ldots \otimes u_k) = \sum_{a_{k+1} \in \mathcal{L}_1 \cap \mathcal{L}_{k+1}} C(u_1,\ldots,u_k,a_{k+1})\cdot a_{k+1},
\end{equation}
where 
\begin{equation}
	C(u_1,\ldots,u_k,a_{k+1}) = \sum_{\phi} \pm e^{2\pi i \int \phi^* \omega} \cdot P e^{\oint \phi^* \beta},
\end{equation}
is a matrix in $\textrm{Hom}(\left.\mathcal{E}_1\right|_{a_{k+1}},\left.\mathcal{E}_{k+1}\right|_{a_{k+1}})$. In the above, the summation is over holomorphic maps $\phi : D^2 \rightarrow \widetilde{M}$ with the following conditions along the boundary: there are $k+1$ points $p_j = e^{2\pi i \alpha_j}$ such that $\phi(p_j) = a_j$ and $\phi(e^{2\pi i \alpha}) \in \mathcal{L}_j$ for $\alpha \in (\alpha_{j-1},\alpha_j)$. Furthermore, $\omega$ is the complexified K\"ahler form and $P$ represents a path-ordered integration, where $\beta$ is the connection of the flat bundle along the local system on the boundary. The functions $C(u_1,\ldots, u_k,a_{k+1})$ encode \textit{open Gromov-Witten invariants} of the Calabi-Yau manifold $\widetilde{M}$ with respect to the Lagrangian branes $\mathcal{L}_i$. This finishes our discussion of the category $\mathcal{F}(\widetilde{M})$. In the case of the elliptic curve, we will restrict the morphisms to be the degree zero pieces of $\textrm{Hom}$ in the above definition and denote the category formed this way by $\mathcal{F}^0(\widetilde{M})$.

Let us next come to the objects which live in the Fukaya category of the elliptic curve $\mathcal{F}^0(\widetilde{E}^{\tau})$. These are special Lagrangian submanifolds which in this case are just lines with non-trivial winding around cycles of $\widetilde{E}^{\tau}$. To define a closed submanifold in $\mathbb{R}^2/(\mathbb{Z} \oplus \mathbb{Z})$ the slope of the line must be rational, and therefore can be given in terms of a pair of integers $(p,q)$. Furthermore, we need to specify the point of intersection with the line $x=0$ (or $y=0$ if $p=0$). We will be considering here only rank $1$ bundles on these Lagrangians, specified by a monodromy around the circle, i.e. a complex phase $\exp(2\pi i \beta)$, $\beta \in \mathbb{R}/\mathbb{Z}$. 

On the derived category side $\mathcal{D}^b(E_{\tau})$ these objects are mapped to line bundles on the mirror torus $E_{\tau} = \mathbb{R}^2/(\mathbb{Z} \oplus \tau \mathbb{Z})$. In the following we will sometimes write $E_{\tau}$ as $E_q$ where $q=e^{2\pi i \tau}$. Recall that a rank $r$ holomorphic vector bundle $F_q(V,A)$ on $E$ is defined by taking the quotient
\begin{equation}
	F_q(V,A) = \mathbb{C}^* \times V / (u,v) \sim (uq,A(u) \cdot v).
\end{equation}
When $V = \mathbb{C}$ and $A = \varphi$ a holomorphic function, we call $L_q(\varphi)$ the line bundle constructed in this way. We define $L \equiv L_q(\varphi_0)$, where $\varphi_0(u) = \exp(-i\pi \tau - 2\pi i z) = q^{-1/2} u^{-1}$. The classical theta function is a section of $L$. For a torus, the theta function depends on the modular parameter $\tau$, the elliptic parameter $z$, and $(c',c'') \in \mathbb{R}^2/\mathbb{Z}^2$, a translation parametrizing different line bundles of the same degree:
\begin{equation}
	\theta[c',c''](\tau,z) = \sum_{m \in \mathbb{Z}} \exp\{2\pi i [\tau(m + c')^2/2 + (m+c')(z+c''))\}.
\end{equation}
In case $(c',c'') = (0,0)$, one simply writes $\theta(\tau,z)$. The $n$ functions $\theta[a/n,0](n\tau,nz), a \in \mathbb{Z}/n\mathbb{Z}$, are the global sections of $L^n$.

Let us now come to the map between the two categories. On the Fukaya side, we consider the Lagrangian branes with slopes
\begin{eqnarray}
	\mathcal{L}_1 & = & (1,0), \nonumber \\
	\mathcal{L}_2 & = & (1,1), \nonumber \\
	\mathcal{L}_3 & = & (1,2).
\end{eqnarray}
These are mapped on the derived category side to line bundles of degrees $0$, $1$, $2$. We have $\mathcal{L}_1 = \mathcal{O}$, which is the sheaf of holomorphic functions. Furthermore, we define $L \equiv \mathcal{L}_2$ and $\mathcal{L}_3 = L^2$. Then the morphisms of the category $\mathcal{F}^0(\widetilde{E}^{\tau})$ are mapped to sections of line bundles as follows
\begin{eqnarray}
	\textrm{Hom}(\mathcal{L}_1,\mathcal{L}_2) & = & H^0(L), \nonumber \\
	\textrm{Hom}(\mathcal{L}_2,\mathcal{L}_3) & = & H^0(L), \nonumber \\
	\textrm{Hom}(\mathcal{L}_1,\mathcal{L}_3) & = & H^0(L^2).
\end{eqnarray}
The product of global sections gives us a map
\begin{equation} \label{eq:thetaproduct}
	m_2 : H^0(L) \otimes H^0(L) \rightarrow H^0(L^2).
\end{equation}
On the side of the Lagrangian branes, the above product formula has to be interpreted in the context of the gluing described by equation (\ref{eq:gluing}). Let us see how this comes about. Note that $\mathcal{L}_1 \cap \mathcal{L}_2 = \{e_1\}$, where we denote by $e_1$ the origin $e_1 \equiv (0,0)$. Also, $\mathcal{L}_2 \cap \mathcal{L}_3 = \{e_1\}$, while $\mathcal{L}_2 \cap \mathcal{L}_3 = \{e_1,e_2\}$ with $e_2 = (1/2,0)$. On the left hand side of (\ref{eq:thetaproduct}), $e_1$ represents the theta function $\theta(\tau,z)$. On the right hand side, the $e_i$ represent a distinguished basis of the two-dimensional space of sections of $L^2$ defined by \cite{Polishchuk:1998db}
\begin{eqnarray}
	e_1 & \leftrightarrow & \theta[0,0](2\tau,2z), \nonumber \\
	e_2 & \leftrightarrow & \theta[1/2,0](2\tau,2z).
\end{eqnarray}
Then (\ref{eq:gluing}) becomes
\begin{equation}
	m_2(e_1 \otimes e_1) = C(e_1,e_1,e_1) \cdot e_1 + C(e_1, e_1, e_2) \cdot e_2,
\end{equation}
where one uses the procedure described above to compute the matrix elements $C$. Thus, $C(e_1,e_1,e_1)$ is given by summing over all triangles with the origin as vertex. In the universal cover of the torus, these correspond to lattice points as vertices and sides of slope $0$, $1$, and $2$. A similar summation holds for $C(e_1,e_1,e_2)$, where the third vertex is the point $(0,1/2)$ up to translation. These summations give
\begin{eqnarray}
	C(e_1,e_1,e_1) & = & \sum_{n=-\infty}^{\infty} \exp[-2\pi A n^2] \nonumber \\
	C(e_1,e_1,e_2) & = & \sum_{n=-\infty}^{\infty} \exp[-2\pi A (n + 1/2)^2].
\end{eqnarray}
We thus obtain
\begin{eqnarray}
	C(e_1,e_1,e_1) & = & \theta[0,0](i2A,0) = \theta[0,0](2\tau,0), \nonumber \\
	C(e_1,e_1,e_2) & = & \theta[1/2,0](i2A,0) = \theta[1/2,0](2\tau,0).
\end{eqnarray}
The product $m_2$ therefore precisely reproduces the addition formula for theta functions
\begin{equation}
	\theta(\tau,z) \theta(\tau,z) = \theta[0,0](2\tau,0)\theta[0,0](2\tau,2z) + \theta[1/2,0](2\tau,0)\theta[1/2,0](2\tau,2z).
\end{equation}
This formula can be easily generalized to cases where $\mathcal{L}_3$ has non-trivial $x$-axis interception $\alpha$ and carries a flat line bundle with non-trivial connection $2\pi i \beta d t_3$ where $t_3$ is a coordinate along $\mathcal{L}_3$. In these cases the $C$ functions have to be replaced by $\theta[\alpha,\beta](2\tau,0)$ and $\theta[1/2+\alpha,\beta](2\tau,0)$. We refer here to \cite{Polishchuk:1998db} for further details.

We now want to close the circle of ideas and connect the present discussion to our discussion of mirror symmetry for the elliptic curve in Section \ref{sec:ellcurve}. The missing link is provided by a nice paper of Zaslow \cite{Zaslow:2005wf} where he uses the above construction to derive the mirror map. Let us see how this comes about. To begin with, we start with the symplectic two-torus $\widetilde{E}^{\tau}$ and construct Lagrangian branes which are mirror dual to sections of line bundles in the torus $E_{\tau}$. Without knowing the mirror map, one can still say that $E_{\tau}$ has a projective embedding as a cubic curve. Since we need three sections of equal degree to realize such an embedding, we learn that the Lagrangian brane $L$ corresponding to the line bundle $\mathcal{O}(1)$ has to have winding number three on the torus $\widetilde{E}^{\tau}$. This can be achieved by taking the base section $S$ corresponding to the sheaf of holomorphic functions $\mathcal{O}$ to be the $x$-axis in the universal cover $\mathbb{R}^2$, and take $L$ to be a line of slope three. Thus we have 
\begin{equation}
	L_k = \{(t,3kt) \textrm{ mod } \mathbb{Z}^2 : t \in \mathbb{R}\}.
\end{equation}
We then define sections (or morphisms on the Fukaya side) $X_i = (i/3,0) \in \textrm{Hom}(S,L)$, $Y_i = (i/6,0) \in \textrm{Hom}(S,L_2)$, and $Z_i = (i/9,0) \in \textrm{Hom}(S,L_3)$, where $i$ is taken mod $3$, $6$, and $9$, respectively. On the derived category side, we have that for example the $X_i$ correspond to sections of the form
\begin{equation}
	X_i = \theta[i/3,0](3\tau,z),
\end{equation}
and similar equations hold for $Y_i$ and $Z_i$. These definitions can be used to work out concrete realizations of the formula (\ref{eq:gluing}), as done in \cite{Zaslow:2005wf}. For example, one has the following relation among the $X_i$ and $Y_i$ coordinates
\begin{equation} \label{eq:XXrel}
	X_i X_j = \sum_{k=0}^l A_{i-j + 3k} Y_{i+j+3k},
\end{equation}
where $A_k := \theta[k/6,0](6\tau,0)$, $k \in \mathbb{Z}/6\mathbb{Z}$. This can be worked out by counting areas of triangles bounded by the points $(i/3,0)$, $(j/3,j)$ and $(\frac{i+j+3k}{6},0)$ on the symplectic torus. A similar formula holds for the product of $Y_i$ with $X_j$:
\begin{equation} \label{eq:XYrel}
	Y_i X_j = \sum_{k=0}^2 B_{2j-i + 6k} Z_{i+j+3k},
\end{equation}
with $B_k = \theta[k/18,0](18\tau,0)$. The final step involves the following observation. If we count all possible monomials which are products of $X_i$ at the cubic level, we arrive at ten independent polynomials:
\begin{equation}
	\{X_0^3,X_1^3,X_2^3,X_0^2X_1, X_1^2 X_2, X_2^2 X_0, X_0^2 X_2, X_1^2 X_0, X_2^2 X_1, X_0 X_1 X_2\}.
\end{equation}
All these products can solely be expressed in therms of the $Z_i$ via the relations (\ref{eq:XXrel}) and (\ref{eq:XYrel}). However, there are only nine $Z_k$, thus we expect a new relation between the $X_k$. Performing the computation one finds \cite{Zaslow:2005wf}
\begin{equation}
	u X_0^3 + u X_1^3 + u X_2^3 + (-2q-p) X_0 X_1 X_2 = 0, 
\end{equation}
where
\begin{eqnarray}
	u = A_2 B_0 + A_1 B_9 \nonumber \\
	p = A_0 B_0 + A_3 B_9 \nonumber \\
	q = A_0 B_6 + A_3 B_3. 
\end{eqnarray}
The modular invariant is then calculated in terms of $\widetilde z = \frac{2q+p}{3u}$\footnote{The coordinate $\widetilde z$ used here is related to the $z$-coordinate from Table \ref{tab:Jtab} through $\widetilde z = 1/3 z^{-1/3}$.}:
\begin{equation}
	j(\tau) = 27 \widetilde z^3 (\widetilde z^3+8)^3(1-\widetilde z^3)^{-3}.
\end{equation}
This is the mirror map as predicted by categorical mirror symmetry. The miracle happens when we expand the right hand side of the above equation. Doing that one obtains the familiar series expansion of the $j$-invariant which gives back the mirror map of Section \ref{sec:ellcurve}!

The author is not aware of a similar derivation of the mirror map for the case of the K3 surface. Obtaining such a representation for that case would be very desirable.

\subsection{ADE string chains and mirror symmetry}
\label{sec:ADE}

It is time to shift gears and focus our attention on Calabi-Yau threefolds which admit elliptic fibrations. For such manifolds, we know that Theorem \ref{maintheo} holds and thus the topological string free energy as well as the partition function admit an expansion in terms of modular forms as outlined in the theorem. Let us look more closely at a particular class of examples. This class consists of non-compact threefolds which are elliptic fibrations over a non-compact base $B$. $B$ is a complex two-dimensional space which is obtained by blowing up an ADE singularity. As such, it has 2-cycles $C^i$ which are $\mathbb{P}^1$'s with negative intersection matrix $\eta^{ij} = - C^i \cdot C^j$ being equal to the Cartan matrix of a simply laced gauge group of ADE type. The structure of the elliptic fiber is such that above each $C^i$ we let it degenerate according to an $I_{N_i}$ Kodaira singularity. To maintain the Calabi-Yau condition the $N_i$ have to be proportional to the Dynkin label of the corresponding node in the ADE Dynkin diagram. 

Topological string partition functions for such geometries were computed in a series of papers \cite{Haghighat:2013gba,Haghighat:2013tka,Gadde:2015tra}. What one finds is the following structure for the topological string partition function $Z^{top}$:
\begin{equation}
	Z^{top}(t_{b,1}, \ldots, t_{b,r},\tau,\vec{t}_{f,i},\lambda) = \sum_{\vec{n}} e^{2\pi i \vec{n} \cdot \vec{t}_b} Z^{\vec{n}}(\tau,\vec{t}_{f,i},\lambda),
\end{equation}
where in the above $\tau$ is the modulus of the elliptic fiber, $t_{b,i}$, $i=1,\ldots,r$ ($r$ is the rank of the ADE Lie algebra here) are the moduli of the $C^i$ curves in the base, and $\vec{t}_{f,i}$ are moduli of the degenerate elliptic fiber above node $i$. Furthermore, one finds that the $Z^{\vec{n}}$ are meromorphic Jacobi forms in the topological string coupling constant $\lambda$ of weight $0$ and index quadratic in the $n_i$. Expanding the $Z^{\vec{n}}$ in powers of $\lambda$ one finds that all coefficients are polynomials in quasi-modular forms in accord with the results of Theorem \ref{maintheo}.

Next, we want to add the affine node to the ADE quiver giving rise to the affine Lie algebra $\widehat{\frak{g}}$. As the base becomes elliptic as well now, the resulting Calabi-Yau threefold, denoted in the following by $X_{N,\widehat{\frak{g}}}$ admits two elliptic fibrations $\pi_1$ and $\pi_2$ which project to fiber and base elliptic curves respectively. In fact, as the degree of degeneration of the elliptic fiber for each node $i$ is different (i.e. depends on the Dynkin label $d_i$), we have the following structure
\begin{equation}
	\pi_{1,i} : X_{N,\widehat{\frak{g}}} \rightarrow \widehat{A}_{d_i N -1}, \quad \pi_2 : X_{N,\widehat{\frak{g}}} \rightarrow \widehat{\frak{g}},
\end{equation}
where by abuse of notation $\widehat{\frak{g}}$ denotes the total space of an elliptic fibration over the unit disc $\mathbb{D}$ such that all fibers are smooth except for the central fiber, where the elliptic curve degenerates to a union of nodal curves of the Kodaira type of $\frak{g}$. $\widehat{A}_{d_i N -1}$ then is the special case where the fiber over the center of the unit disc degenerates according to Kodaira type $I_{d_i N}$. The Calabi-Yau manifolds $X_{N,\widehat{\frak{g}}}$ are important in string theory for F-theory constructions of so called \textit{little string} theories. In the following we shall denote the modulus of the base elliptic curve by $\rho$.

In the following, we want to look at the mirror Calabi-Yau manifold of $X_{N, \widehat{\frak{g}}}$ which was constructed in \cite{Haghighat:2017vch} (see also \cite{Kanazawa:2016tnt} for an earlier result where $\widehat{\frak{g}} = \widehat{A}_r$) and is given by the following equation
\begin{equation}
	u v = F^{\textrm{open}}(z_1,z_2),
\end{equation}
where $F^{\textrm{open}}$ is the open Gromov-Witten potential. The equation 
\begin{equation}
	F^{\textrm{open}} = 0 
\end{equation}
is then known as the so called \textit{mirror curve}. The mirror curve contains all the relevant information about the mirror Calabi-Yau manifold and in order to learn more about it, we want to focus in the following on a particular example where the above equation is known very explicitly. The example is given by taking $\frak{g}$ to be $A_{N-1}$. In this case one arrives at \cite{Kanazawa:2016tnt,Haghighat:2017vch}
\begin{equation} \label{eq:mirrorcurve}
	F^{\textrm{open}} = \sum_{i=0}^r \sum_{l=0}^{N-1} K_{i,l} \Delta_{i,l} \Theta_2\left[\begin{array}{c}(\frac{i}{r+1},\frac{l}{N})\\(\frac{-(r+1)\rho}{2},-\frac{N\tau}{2})\end{array}\right]\left((r+1)z_1,N z_2; \left[\begin{array}{cc}(r+1)\rho & \sigma \\ \sigma & N \tau\end{array}\right]\right).
\end{equation}
Let us explain the notation. $\Delta_{i,l}$ are open Gromov-Witten generating functions and $K_{i,l}$ are given by
\begin{equation}
	K_{i,l} = q_{\rho}^{\frac{i}{2}-\frac{i^2}{2(r+1)}} q_{\tau}^{\frac{l}{2}-\frac{l^2}{2N}}.
\end{equation}
The theta function in (\ref{eq:mirrorcurve}) is the genus $2$ theta function. The genus $g$ theta function is defined as follows
\begin{equation}
	\Theta_g\left[\begin{array}{c}
	\vec{a}\\
	\vec{b}
	\end{array}\right](\vec{z};\Omega) = \sum_{\vec{n} \in \mathbb{Z}^g} \exp\left(\frac{1}{2}(\vec{n} + \vec{a})^t \Omega (\vec{n}+\vec{a}) + (\vec{n} + \vec{a})\cdot (\vec{z} + \vec{b})\right),
\end{equation}
where $\Omega$ is an element of the Siegel upper half plane
\begin{equation}
	\mathbb{H}_g = \{ \Omega \in M_g(\mathbb{C})~ | ~\Omega^t = \Omega, ~\textrm{Im}(\Omega) > 0\}.
\end{equation}
Equation (\ref{eq:mirrorcurve}) defines a conic fibration over the abelian surface $\mathbb{C}^2/(\mathbb{Z}^2 \oplus \Omega \mathbb{Z}^2)$ with discriminant being the genus $(r+1)N +1$ curve $F^{\textrm{open}} = 0$, and $u$ and $v$ are sections of suitable line bundles over the abelian surface. 

Let us attempt to interpret the thus specified equation for the mirror curve in the light of categorical mirror symmetry. First of all, we note that the $\Delta_{i,l}$ can be interpreted in terms the functions $m_k$ defined in Section \ref{sec:newellcurve}, namely they are disc instanton gnerating functions for certain configurations of Lagrangian branes. Now what are these Lagrangian branes? The difference between the situation here and the one in the previous section is that now we are effectively dealing with a doubly elliptic surface. Thus Lagrangian lines become Lagrangian planes and morphisms become sections of line bundles on an abelian surface. Such sections are given in terms genus $2$ theta functions, hence their appearance. In fact, in direct analogy to Section \ref{sec:newellcurve} the $\Theta_2$ summands correspond to the $Z_i$ appearing in the equation for the mirror elliptic curve. An interesting open question is whether this analogy can be carried further by constructing the analogues of the sections $X_i$ in the last section and deriving the full mirror map for the present case. One way to attack this problem would be to restrict the section of the abelian surface given by $F^{\textrm{open}}$ to the elliptic curve $E_{\rho}$  and derive the mirror map for the restricted section. In fact, this was the path taken in \cite{Haghighat:2017vch} for deriving (\ref{eq:mirrorcurve}).

\section*{Acknowledgment}

I would like to thank the organizers of the event ``Perspectives of Mathematics in the $21^{\textrm{st}}$ Century: Conference in Celebration of the $90^{\textrm{th}}$ Anniversary  of Mathematics Department of Tsinghua University" where some of the results of this article were introduced to a wider audience for the first time. Furthermore, many thanks go to Rui Sun who helped with literature search and many clarifying discussions.

\end{document}